\documentclass[twocolumn,aps,prl,groupedaddress,showpacs]{revtex4}
\usepackage{amssymb}
\usepackage{amsmath}
\usepackage{bm}
\usepackage{graphicx}
\usepackage{color}

\input{tcilatex}

\begin{document}

\title{Long ranged singlet proximity effect in ferromagnetic nanowires}
\author{F. Konschelle$^{(1)}$}
\affiliation{$^{(1)}$ Kavli Institute of Nanoscience, Delft University of Technology,
P.O. Box 5046, 2600 GA Delft, The Netherlands}
\author{J. Cayssol$^{(2)}$}
\author{A. Buzdin$^{(2)}$}
\altaffiliation{\emph{Also at }Institut Universitaire de France.}
\affiliation{$^{(2)}$Universit\'{e} de Bordeaux, CPMOH, UMR 5798, 33405 Talence, France}
\date{\today}

\begin{abstract}
Recently a long ranged superconductor/ferromagnet (S/F) proximity effect has
been reported in Co crystalline nanowires \cite[Nature, \textbf{6} 389 (2010)%
]{Co wire}. Since the authors of \cite{Co wire} take care to avoid the
existence of magnetic domains, the triplet character of the long ranged
proximity effect is improbable. Here we demonstrate that in the
one-dimensional ballistic regime the standard singlet S/F proximity effect
becomes long ranged. We provide an exact solution for the decay of the
superconducting correlations near critical temperature ($T_{c}$) and for
arbitrary impurities concentration. In particular we find a specific regime,
between the diffusive and ballistic ones, where the decay length is simply
the electronic mean-free path. Finally possible experiments which could
permit to elucidate the nature of the observed long ranged proximity effect
in Co nanowires are discussed.
\end{abstract}

\maketitle

It is well known that the superconducting proximity effect in a diffusive
ferromagnetic (F) metal is rather short-ranged due to the large value of the
ferromagnetic exchange field $E_{ex}\sim \left( 500-5000\right) K$ acting on
the electron spins. In the ferromagnet, the singlet Cooper pair wave
function experiences an oscillatory decay whose characteristic length $\xi
_{f}=\sqrt{\hslash D/E_{ex}}$ enters the nanoscopic range for typical values
of the diffusion constant $D$ and exchange field in the ferromagnet \cite%
{BuzdinRMP2005,GolubovRMP2004}. This is in a sharp contrast with the
corresponding decaying length in normal metals $\xi _{n}=\sqrt{\hslash
D/k_{B}T}\gg \xi _{f}$ which can reach microns at low enough temperatures.

The presence of the non-uniform magnetization may strongly modify the S/F
proximity effect for singlet superconductors, and in the diffusive regime
the induced triplet correlations can penetrate at the large distance $\xi
_{n}$ \cite{BergeretRMP2005}. In the clean limit and in the presence of
domains, the triplet proximity effect also increases the decay length up to
the thermal decay length $\xi _{nb}=\hslash v_{F}/k_{B}T$ \cite{HouzetBuz},
where $v_{F}$ is the Fermi velocity. The long range proximity effect has
been observed in Ho \cite{Sosnin} and in the half metal CrO$_{2}$ \cite{Xiao}%
. Probably its origin is related with the presence of magnetic domains in
Ho, and with magnetic scattering at the S/F interface in the case of CrO$%
_{2} $. Recently the controlled transition between the short ranged singlet
and the long ranged triplet proximity effect has been reported in Co based
Josephson junctions using Co/Ru/Co \cite{BirgePRL} and Ho/Co/Ho \cite%
{Robinson2010} as magnetic barriers, respectively. Previously the $200$ nm
decay length for the coherence effect was observed in a system with
mesoscopic Co wire \cite{Giroud}. The geometry of the Co electrodes in the
setup of \cite{Giroud} was nevertheless 2D-like and probably the observed
long ranged proximity effect was related with the presence of the magnetic
domains which generated the triplet superconducting correlations \cite%
{BergeretRMP2005}.

Very recently the Josephson current through the Co crystalline nanowires as
long as $600$ nm was reported \cite{Co wire}. Interestingly the experiments
on the Co crystalline nanowires \cite{Co wire} were performed after applying
a strong magnetic field which makes improbable the existence of the magnetic
domains. The diameters of the Co nanowires \cite{Co wire} \ were $40$ nm and 
$80$ nm and the proximity effect was substantially weaker for thicker
nanowires. Note that in the S/F/S structures with Co film as a F layer, the
long ranged proximity effect is absent and the characteristic decay length
of superconductivity is around $5$ nm \cite{Robinson,Birge}. Therefore the
nanoscopic character of the Co wire seems to be crucial for the observation
of the long ranged proximity effect.

In this Rapid Communication we present a theoretical analysis of the S/F
proximity effect in the 1D case taking in mind a thin Co nanowire. For this
simple model it is possible to obtain an exact solution of Eilenberger
equations with arbitrary impurity scattering. The decay of the
superconducting correlations is governed by the electron mean free path $l$.
We believe that for the Co nanowire $l$ may be in the range $(50-100)$ nm
which could explain the long ranged proximity effect observed in \cite{Co
wire}. We also discuss the alternative scenarios including dead layers or
triplet correlations. Finally we suggest future experiments to discriminate
between those different types of explanations.

The diffusive regime for the proximity effect in ferromagnets corresponds to
the limit $E_{ex}\tau \ll 1$ \cite{Baladie} which could be realized in weak
ferromagnets with CuNi or PdNi alloys as F layer \cite{Ryazanov1,Kontos} but
not in the systems with strong ferromagnet like Co, Ni or Fe \cite%
{Robinson,Birge}. However in ballistic regime the S/F proximity effect is
substantially different. The Cooper pair wave function oscillates inside the
ferromagnet at the characteristic length $\xi _{fb}=\hslash v_{F}/E_{ex}$ 
\cite{BuzBul} but it decays much slowly : in 3D system as $\sim \left( \xi
_{fb}/x\right) $, where $x$ is the distance from the S/F interface, in 2D
systems as $\sim \sqrt{\left( \xi _{fb}/x\right) },$ and no decay in 1D case 
\cite{Konschelle,Cayssol2004}. Sure this statement is true for the distances 
$x$ smaller than thermal decay length $\xi _{nb}=\hslash v_{F}/k_{B}T$ and
mean free path $l=v_{F}\tau $, where $\tau $ is the average time between
impurity scattering events. Otherwise the range of the proximity effect will
be fixed by the smallest length \cite{Bul,Efet,Linder}.

Near the superconducting critical temperature, the Eilenberger equations 
\cite{Eilenberger1968} may be linearized and they read for the 1D case: 
\begin{equation}
\left( \pm \dfrac{v_{F}}{2}\dfrac{\partial }{\partial x}+\omega +iE_{ex}+%
\dfrac{1}{2\tau }\right) f_{\pm }=\dfrac{1}{4\tau }(f_{+}+f_{-}),
\label{eilenberger}
\end{equation}%
where $f_{+}$ and $f_{-}$ are the anomalous Eilenberger Green functions for $%
v_{F}>0$ and $v_{F}<0$ respectively and $\omega $ is the fermionic Matsubara
frequency. Note that $(f_{+}+f_{-})/2$ is simply the average over the Fermi
surface in the 1D case.

After introducing the symmetric and antisymmetric combinations $%
F_{s}=(f_{+}+f_{-})/2$ and $F_{a}=(f_{+}-f_{-})/2$, one obtains that $F_{s}$
obeys the following second order differential equation:%
\begin{equation}
\left( \omega +iE_{ex}\right) F_{s}-\frac{v_{F}^{2}}{4\left( \omega +iE_{ex}+%
\dfrac{1}{2\tau }\right) }\frac{\partial ^{2}F_{s}}{\partial x^{2}}=0,
\label{usadel}
\end{equation}%
while the antisymmetric combination can be obtained from 
\begin{equation}
F_{a}=-\frac{v_{F}}{2}\left( \omega +iE_{ex}+\frac{1}{2\tau }\right) ^{-1}%
\frac{\partial F_{s}}{\partial x}.  \label{Fa}
\end{equation}

Let us now sketch two limiting cases of Eq.$\left( \ref{usadel}\right) $. In
the diffusive limit $E_{ex}\tau \ll \hslash $, Eq.(\ref{usadel}) becomes:%
\begin{equation}
\left( \omega +iE_{ex}\right) F_{s}-\frac{v_{F}^{2}\tau }{2}\frac{\partial
^{2}F_{s}}{\partial x^{2}}=0  \label{dirty}
\end{equation}%
which is the Usadel equation with the diffusion coefficient $D=v_{F}^{2}\tau 
$. In the ballistic limit $E_{ex}\tau \gg \hslash $, Eq.(\ref{usadel})
becomes:%
\begin{equation}
\left( \omega +iE_{ex}\right) ^{2}F_{s}-\frac{v_{F}^{2}}{4}\frac{\partial
^{2}F_{s}}{\partial x^{2}}=0.  \label{clean}
\end{equation}%
which contains the same information as the Eilenberger equation Eq.(\ref%
{eilenberger}) in the absence of disorder ($1/\tau=0$).

Solving the two previous equations (\ref{dirty},\ref{clean}) for $E_{ex}\gg
T_{c}$ shows that in the diffusive regime the characteristic oscillatory and
damped behavior of $F_{s}$ occur with the same characteristic length $\xi
_{f}$, whereas the ballistic regime may exhibit no attenuation at all. In
the following, we investigate the effect of a small amount of impurities
into the ferromagnetic wire. We show that this regime is characterized by
two distinct lengths: the ballistic ferromagnetic oscillation one $\xi _{fb}$
and the damped one $l$ induced by the impurities into the wire.

The general solution of Eq.(\ref{usadel}) for arbitrary impurity scattering
is simply:%
\begin{equation}
F_{s}=Ae^{qx}+Be^{-qx},
\end{equation}%
with 
\begin{equation}
q^{2}=\frac{4}{v_{F}^{2}}\left( \omega +iE_{ex}+\frac{1}{2\tau }\right)
\left( \omega +iE_{ex}\right) .
\end{equation}%
Using Eq.(\ref{Fa}) and the approximation $E_{ex}\tau \gg 1$, one obtains
the antisymmetric anomalous propagator: 
\begin{equation}
F_{a}=-Ae^{qx}+Be^{-qx}.
\end{equation}%
Assuming superconducting electrodes much thicker than the F nanowire, we
apply the rigid boundary condition $f_{+}(x=-L/2)=\Delta e^{-i\varphi
/2}/\omega $. Then we find the Eilenberger propagator for the
superconducting correlations originating from the left electrode: 
\begin{equation}
f_{+}=\frac{\Delta }{\omega }e^{-i\varphi /2}e^{-q(x+L/2)},
\end{equation}%
where we suppose the temperature close to $T_{c}$. The similar condition $%
f_{-}(x=L/2)=\Delta e^{-i\varphi /2}/\omega $ yields the correlations from
the right superconductor:%
\begin{equation}
f_{-}=\frac{\Delta }{\omega }e^{i\varphi /2}e^{q(x-L/2)}.
\end{equation}%
We now evaluate the Josephson supercurrent for this single channel situation:%
\begin{eqnarray}
I &=&\pi e\nu _{0}T\sum\limits_{\omega }v_{F}Im(f_{-}f_{-}-f_{+}f_{+}) 
\notag \\
&=&2\pi e\nu _{0}v_{F}T\sum\limits_{\omega }\left( \frac{\Delta }{\omega }%
\right) ^{2}\cos \left( \frac{2E_{ex}L}{v_{F}}\right) e^{-L/2l}\sin \varphi 
\notag \\
&=&\dfrac{\pi }{2}e\nu _{0}v_{F}\dfrac{\Delta ^{2}}{T_{c}}\cos \left( \frac{%
2E_{ex}L}{v_{F}}\right) e^{-L/2l}\sin \varphi ,  \label{quasi-clean}
\end{eqnarray}%
where $\nu _{0}=k_{F}/\pi /E_{F}=(E_{F}a)^{-1}$ is the one-dimensional
density of states per unit length (including 2 spin directions).

In the pure limit, $l\gg L$, the critical current exhibits undamped
oscillations: 
\begin{eqnarray}
I_{c0} &=&2\pi e\nu _{0}v_{F}T\sum\limits_{\omega }\left( \frac{\Delta }{%
\omega }\right) ^{2}\cos \left( \frac{2E_{ex}L}{v}\right)  \notag \\
&=&\dfrac{\pi }{2}e\nu _{0}v_{F}\dfrac{\Delta ^{2}}{T_{c}}\cos \left( \frac{%
2E_{ex}L}{v_{F}}\right)
\end{eqnarray}%
whose amplitude gives an estimate for the single channel critical current.
The above result was derived near the critical temperature. Concerning order
of magnitudes, it can be safely extrapolated to low temperature by
substituting $\Delta =\Delta (T)$ by $\Delta (T=0)\simeq T_{c}$. We
introduce the quantum of resistance $R_{N}=h/e^{2}$ and find: 
\begin{equation}
I_{c0}\sim \dfrac{\pi }{2}e\nu _{0}v_{F}\dfrac{\Delta ^{2}}{T_{c}}\text{ }%
\sim \frac{T_{c}}{eR_{N}}.
\end{equation}%
Hence we can estimate for the experimental conditions \cite{Co wire}: 
\begin{equation}
eR_{N}I_{c0}\sim 1\text{ mV \ and \ so \ }I_{c0}\sim 40\text{ nA.}
\end{equation}%
This maximal current per mode is damped by the exponential factor $e^{-L/2l}$%
.

For a metallic nanowire with a cross section $S$, the number $M$ of
transverse channels typically amounts $S/a^{2}\simeq (d/a)^{2}\simeq \left(
10^{4}-10^{5}\right) $. This relatively large number of channels may
compensate the exponential suppression of the critical current which can be
evaluated as $I_{c}=Me^{-L/2l}I_{c0}$. In experiment \cite{Co wire}, the
critical current exceeds $10$ $\mu $A and $M\simeq 10^{5}$ ($d=\left(
40-80\right) $ nm, $L=600$ nm and $a=0.1$ nm) thereby requiring a rather
long mean free path, namely $l\geq 60$ nm. On the basis of the Drude model 
\cite{Co wire}, the estimate of the mean free path\textbf{\ }$l$ is quite
sample dependent with values ranging in the window $l$ $\sim (2-10)$ nm. In
our approach such short mean free paths would lead to a very strong
suppression of the current. Nevertheless we might question if those values $%
l $ $\sim (2-10)$ nm represent the genuine mean free path inside the
ferromagnet. Indeed the authors \cite{Co wire} note an important diffusion
of the W atoms inside Co nanowires. Therefore the region near the W/Co
interface may contribute to the measured resistivity and the genuine or
intrinsic mean-free path $l$ (defined between the contact regions) of the Co
nanowire may be substantially higher than the simple Drude estimate.

Long range triplet proximity effect, caused by some magnetic inhomogeneity,
is also a possible scenario to explain the experiment \cite{Co wire}. A
novel kind of Josephson junctions S/X$_{1}/$F/X$_{2}/$S with 3 magnetic
layers (X$_{1}$, F, X$_{2}$) has been proposed \cite{HouzetBuz} and
successfully implemented in a controlled way \cite{BirgePRL,Robinson2010}. A
central region (F) is used to suppress singlet Josephson current while the
side domains create triplet correlations. In the original theoretical
set-up, the X$_{1}$, F and X$_{2}$ were three monodomains with noncollinear
magnetizations. In the recent experiment \cite{BirgePRL}, X$_{1}$ and X$_{2}$
were layers of weak ferromagnets while F was itself a Co bilayer, Co/Ru/Co.
In Ref.\cite{Robinson2010} the synthetic Ho/Co/Ho ferromagnetic trilayer was
used to the same purpose. More specifically it was predicted \cite{HouzetBuz}
and confirmed experimentally \cite{BirgePRL} that the production of triplet
correlations is maximal for an optimal size of the X$_{1}$ and X$_{2}$
layers, which is on the order of $\xi _{fb}$. The authors of \cite{Co wire}
mention that though the Co nanowires must be single domain, the
inhomogeneous magnetic moments may be produced in the W-Co contact regions.
Following \cite{HouzetBuz} and \cite{BirgePRL,Robinson2010}, the size of
such regions must be of order $\xi _{f}$ to maximize the triplet long range
effect. Nevertheless this contact region is likely to be\ of the atomic size 
$a$ and in such a case the triplet Josephson current would be reduced by a
factor $(a/\xi _{fb})^{2}\simeq \left( 10^{-2}-10^{-1}\right) $. The optimal
triplet supercurrent is $M\simeq 10^{5}$ times the single channel S/N/S
supercurrent ($0.04$ $\mu $A). Hence our estimation of the triplet effect in 
\cite{Co wire} is $I_{c}=\left( 4-40\right) $ $\mu $A which is still in
agreement with the reported supercurrents.

In order to discriminate between the two scenarios described above, it is
crucial to know whether the Co nanowires are in the diffusive or in the
ballistic limit, and also to have an experimental determination of the
Josephson current-phase relation \cite{BuzdinRMP2005,GolubovRMP2004}. Indeed
the presence or absence of a second harmonic at low temperature allows a
distinction between those two scenarios. In the singlet correlation
scenario, the second harmonic will be strongly suppressed by disorder (by an
additional factor $e^{-L/2l}$ relative to the first harmonic), while for
triplet proximity effect the second harmonic is expected to be of the same
order of magnitude as the first one, at low temperature.

Besides, it would also be interesting to know if the S/F/S structures with
Co nanowire may be in the $\pi $ state. This can be determined by realizing
a SQUID with the Co nanowire Josephson junction as one arm and a standard
Josephson junction on the other arm.

Note that a third possible scenario might involve a dead layer at the
surface of the Co nanowire, where magnetism is strongly suppresses thereby
leading to the proximity effect reported in \cite{Co wire}. Indeed the Co
wires are surrounded by an insulating oxide coating and a dead layer
(without ferromagnetic order) might appear between this insulating layer and
the Co wire core. The presence of such dead layer was already signaled in Co
films, with typical size $1$ nm \cite{Robinson}. In the extreme case, the
magnetism is completely absent within this layer and the corresponding
supercurrent would not have the exponential decay. In such a case the number
of channels would be $M_{surf}=d/a\simeq 10^{3}$ and the critical current
would reach quite large values $10^{2}\mu $A, even for long nanowires. Note
that in presence of such a deadlayer and at low temperature, the second
harmonic of the current-phase relation would be as large as the first
harmonic, similar to the case of the triplet long ranged proximity effect.

In conclusion, we have found that the long ranged S/F proximity effect
reported recently in \cite{Co wire} can be explained by three distinct
scenarios. In the first scenario, the Josephson current is associated with
weakly damped singlet superconducting correlations. In the second scenario,
the contact regions produce triplet correlations thereby leading to long
ranged proximity effect. Finally in a third scenario a non magnetic
deadlayer provides a channel for long ranged propagation of the
superconducting correlations. We further suggest to realize the
low-temperature determination of the current-phase relation to discriminate
between those scenarios.

This work was supported by the Agence Nationale de la Recherche Grant No.
ANR-07-NANO-011-05: ELEC-EPR.

\bigskip

\end{document}